\documentclass[reprint,superscriptaddress,aps,prb,twocolumn,floatfix]{revtex4-2}
\usepackage{setspace}
\usepackage{amsmath}
\usepackage{bm}
\usepackage{graphicx}
\usepackage{verbatim}
\usepackage{amsfonts}
\usepackage{amssymb}
\usepackage{textcomp}
\usepackage{mathrsfs}
\usepackage{mathtools}
\usepackage{url}
\usepackage{caption}

\usepackage{xcolor}
\usepackage[colorlinks,linkcolor=blue,anchorcolor=blue,citecolor=blue,urlcolor=black]{hyperref}
\usepackage[hyphenbreaks]{breakurl}
\usepackage{ragged2e}
\usepackage{float}
\usepackage{lineno}
\DeclareGraphicsExtensions{.pdf,.eps,.png,.jpg,.mps}
\begin{document}

\title{Velocity-comb modulation transfer spectroscopy}
\author{Xiaolei Guan$^{1}$, Zheng Xiao$^{1}$, Zijie Liu$^{1}$, Zhiyang Wang$^{1}$, Jia Zhang$^{1}$, Xun Gao$^{1}$,\\ Pengyuan Chang$^{2}$, Tiantian Shi$^{3,\dagger}$, and Jingbiao Chen$^{1,4}$ \\
\vspace{3pt}
$^1$State Key Laboratory of Advanced Optical Communication Systems and Networks,\\ Institute of Quantum Electronics, School of Electronics, Peking University, Beijing 100871, China\\
$^2$Institute of Quantum Information and Technology,\\ Nanjing University of Posts and Telecommunications, Nanjing 210003, China\\
$^3$National Key Laboratory of Advanced Micro and Nano Manufacture Technology, School of Integrated Circuits, Peking University, Beijing 100871, China\\
$^4$Hefei National Laboratory, Hefei 230088, China\\
\vspace{3pt}
Corresponding authors: $^\dagger$tts@pku.edu.cn}



\date{\today}

\maketitle
\noindent
\large\textbf{Abstract} \\
\normalsize\textbf{
Sub-Doppler laser spectroscopy is a crucial technique for laser frequency stabilization, playing a significant role in atomic physics, precision measurement, and quantum communication. However, recent efforts to improve frequency stability appear to have reached a bottleneck, as they primarily focus on external technical approaches while neglecting the fundamental issue of low atomic utilization ($\textless$ 1\%), caused by only near-zero transverse velocity atoms involved in the transition. Here, we propose a velocity-comb modulation transfer spectroscopy (MTS) solution that takes advantage of the velocity-selective resonance effect of multi-frequency comb lasers to enhance the utilization of non-zero-velocity atoms. In the probe-pump configuration, each pair of counter-propagating lasers interacts with atoms from different transverse velocity-comb groups, independently contributing to the spectral amplitude and signal-to-noise ratio. Preliminary proof-of-principle results show that the frequency stability of the triple-frequency laser is optimized by nearly a factor of $\sqrt{3}$ compared to the single-frequency laser, consistent with theoretical expectations. With more frequency comb components, MTS-stabilized lasers are expected to achieve order-of-magnitude breakthroughs in frequency stability, taking an important step toward next-generation compact optical clocks. This unique method can also be widely applied to any quantum system with a wide velocity distribution, inspiring innovative advances in numerous fields with a fresh perspective.
}

\vspace{3pt}
\maketitle
\noindent
\large\textbf{Introduction} \\
\normalsize
\noindent Laser spectroscopy \cite{1} has been one of the most active fundamental research fields in optics since the advent of lasers. It fully exploits the unique properties of lasers, overcoming the limitations of traditional spectroscopy in addressing key challenges. In 1971, Arthur L. Schawlow et al. proposed the saturated absorption spectroscopy (SAS) method \cite{2}, which successfully eliminated the spectral broadening induced by the Doppler effect \cite{3,4,5,6,7,8,9}. Since then, sub-Doppler laser spectroscopy has advanced rapidly, driving the development of various high-resolution spectroscopic techniques, including polarization spectroscopy (PS) \cite{10,11,12,29}, two-photon spectroscopy (TPS) \cite{13,14,15,16,17,31}, and modulation transfer spectroscopy (MTS) \cite{18,19,20,21,22}. These techniques are now widely applied in precision measurement \cite{23,24,25}, chemical analysis \cite{66}, and biomedical fields \cite{67}.

A typical application of sub-Doppler laser spectroscopy is laser frequency stabilization. Unlike high-finesse optical cavities \cite{26,27,28}, it enables highly stable laser outputs directly corresponding to quantum transition frequencies. Currently, iodine-based MTS has demonstrated the best performance, achieving frequency stability on the order of $10^{-15}$ with good reproducibility \cite{32,33}, making it recommended as a wavelength standard. MTS-stabilized lasers using alkali metal atoms have also reached the $10^{-14}$ level \cite{34,21,69,65,68}. However, the existing configuration, in which single-frequency probe and pump beams propagate in opposite directions through the vapor cell, allows only a small fraction of atoms (or molecules) with near-zero transverse velocities to participate in the transition. The effective utilization of atoms is insufficient ($\textless$ 1\%), posing significant challenges to further improving the signal-to-noise ratio ($S/N$) of the spectrum. According to the Allan deviation formula $\sigma_y(\tau)=\frac{1}{K}\frac{\Delta\nu}{\nu_0 S/N \sqrt{\tau}}$, the $S/N$ of the quantum reference spectrum is one of the key factors determining laser frequency stability. Therefore, improving the utilization ratio of atoms could be a promising method for the future development of MTS stabilization and other sub-Doppler laser spectroscopy-based stabilization techniques.

\begin{figure*}[htbp]
\centering
\captionsetup{singlelinecheck=no, justification = RaggedRight}
\includegraphics[width=\textwidth]{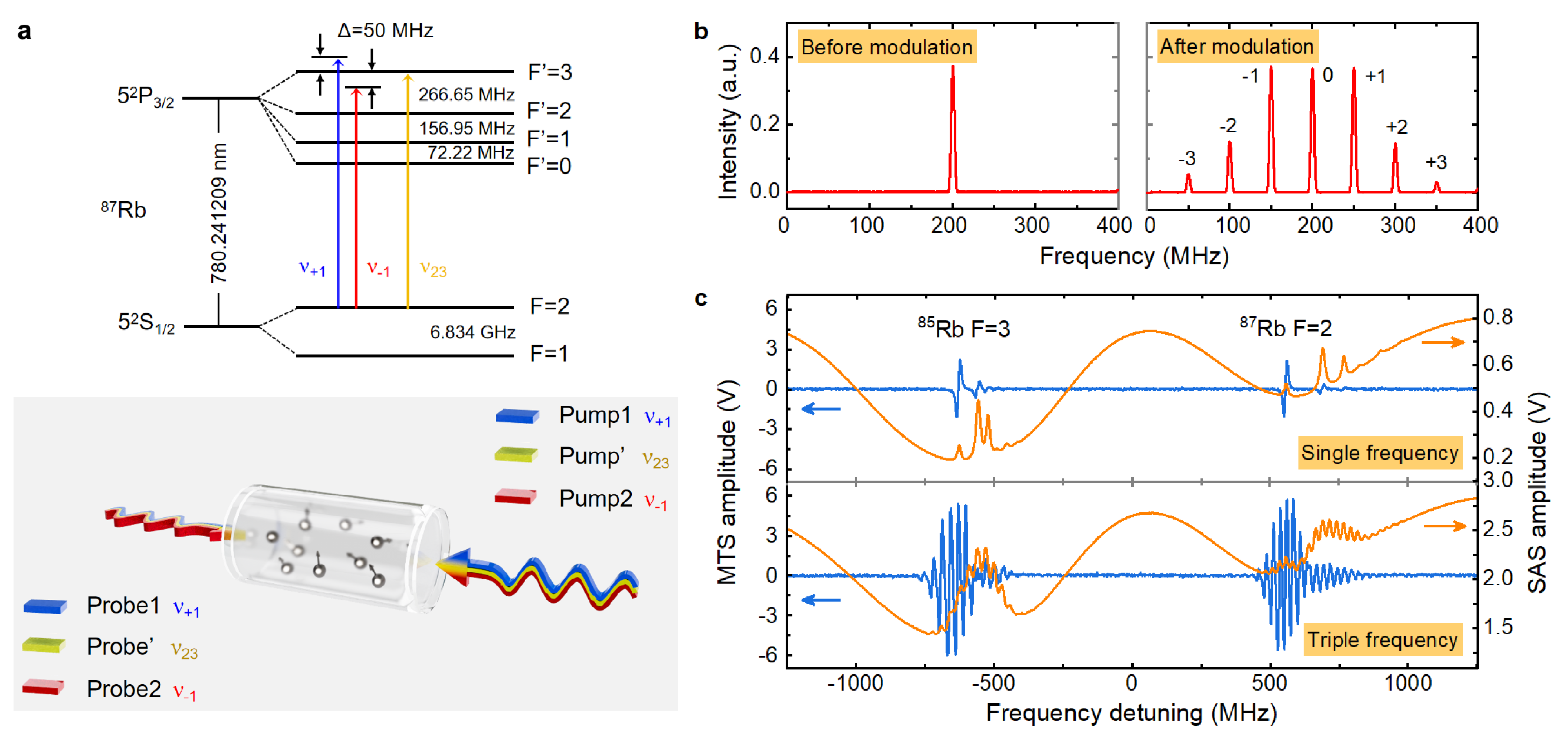}
\caption{\textbf{Working principle of the velocity-comb MTS.} \textbf{a} Relevant $^{87}$Rb atomic energy levels and the interaction process of non-zero velocity atoms with multi-frequency comb lasers. The yellow component ($\nu_{23}$) represents the frequency of the $F$=2 $\rightarrow$ $F’$=3 transition, while the blue ($\nu_{+1}$) and red ($\nu_{-1}$) components correspond to the frequencies of the +1 and -1 order sidebands generated by modulation, respectively. The modulation frequency ($\Delta$) of the electro-optic modulator (EOM1) is 50 MHz. In the thermal Rb vapor cell, atoms with various velocity groups are present, and their overall velocity distribution follows the Maxwell-Boltzmann distribution. \textbf{b} Spectral distribution of the laser before and after modulation by EOM1. Here, the center frequency of 200 MHz represents the frequency shift prior to heterodyne measurements. \textbf{c} Comparison of the SAS signal (orange curve) and MTS signal (blue curve) for the single-frequency and triple-frequency lasers. The left and right sides correspond to the $^{85}$Rb 5$^2$S$_{1/2}$ $F$=3 and $^{87}$Rb 5$^2$S$_{1/2}$ $F$=2 to 5$^2$P$_{3/2}$ transitions, respectively.   
}
\label{fig1}
\end{figure*}

In 2022, H. Shang et al. \cite{35} from our group first proposed the concept of ``velocity-grating'' atom interferometry and conducted theoretical studies using a thermal calcium beam as the frequency reference. The results demonstrated that this method could enhance the amplitude of optical Ramsey fringes by a factor of 1000 or more. In 2024, W. G. Tobias et al. \cite{36} experimentally verified the feasibility of this method in a laser-cooled $^{40}$Ca beam optical clock. Their results showed that the Allan deviations for single-frequency, triple-frequency, and multi-frequency spectroscopy were (31.4, 16.5, 3.4) $\times$ $10^{-15}$/$\sqrt{\tau}$, respectively. In this work, we focus on the thermal vapor-cell system and propose a velocity-comb MTS frequency stabilization scheme, which similarly tackles the challenge of improving atomic utilization by leveraging the velocity-selective resonance effect of multi-frequency comb lasers. Considering that iodine-based MTS requires long vapor cells and frequency-doubled lasers, we choose alkali metal Rb atoms as the frequency reference. The multi-frequency comb components with equal frequency intervals are generated by external phase modulation of a single-frequency laser operating at 780 nm. When applied to the atomic vapor cell in the MTS configuration, each pair of counter-propagating probe and pump lasers interacts with atoms from different transverse velocity groups, independently contributing to the spectral amplitude and $S/N$. Since these atoms exhibit a comb-like distribution in the transverse velocity domain, we refer to the resulting high-resolution spectrum as ``velocity-comb'' MTS. Specifically, using a triple-frequency laser as an example, we compare its Allan deviation with that of a conventional single-frequency laser stabilized by MTS, obtaining values of 4.3 $\times$ $10^{-12}$/$\sqrt{\tau}$ and 7.5 $\times$ $10^{-12}$/$\sqrt{\tau}$ (for $\tau$ from 1 s to 10 s), respectively. This result validates the effectiveness of our scheme in enhancing laser frequency stability.

As a preliminary proof of principle, this study does not yet incorporate more frequency components or fully address environmental influences. Future work, such as the implementation of cascaded modulation \cite{50,51} or the use of an optical frequency comb \cite{52} as the frequency source, is expected to further improve the frequency stability of MTS-stabilized lasers by orders of magnitude. Moreover, this scheme is universal for any quantum system with a wide velocity distribution, showing great potential in the realization of $10^{-16}$-level iodine molecular optical frequency standards \cite{19,33}, as well as offering a new pathway to improve the $S/N$ of some weakly absorbing frequency standards (such as methane \cite{37,38} and carbon dioxide \cite{39}). More importantly, it could serve as a complementary technique to existing sub-Doppler laser spectroscopy, enriching current textbooks \cite{1} and laboratory experiments, thus inspiring more interesting connections in fundamental scientific research.

\vspace{6pt}
\noindent
\large\textbf{Results}\\
\noindent
\normalsize
\textbf{Experimental principle}\\
\noindent
Figure \ref{fig1}a illustrates the energy level scheme and working mechanism of the velocity-comb MTS. A homemade 780 nm external cavity diode laser (ECDL) with single-frequency output is phase-modulated by an electro-optic modulator (EOM1) with a resonance frequency of 50 MHz, generating multiple sidebands in addition to the main frequency (details of the experimental setup are available in the \hyperref[Methods]{Methods} section). Focusing on the $\pm$1st-order sidebands as a representative case, we analyze the spectral amplitude enhancement principle based on the $^{87}$Rb 5$^2$S$_{1/2}$ $F$=2 $\rightarrow$ 5$^2$P$_{3/2}$ $F’$=3 transition.

As shown in Fig. \ref{fig1}a, $\nu_{23}$ represents the frequency of the $F$=2 $\rightarrow$ $F’$=3 transition. $\nu_{+1}=\nu_{23}+\Delta$ and $\nu_{-1}=\nu_{23}-\Delta$ represent the frequencies of the +1 and -1 order sidebands generated by EOM1 modulation, respectively, where $\Delta$ = 50 MHz is the modulation frequency of EOM1. When the laser frequency is scanned to the $F$=2 $\rightarrow$ $F’$=3 transition, the three frequency components, $\nu_{23}$, $\nu_{+1}$, and $\nu_{-1}$, simultaneously interact with the atomic medium as both probe and pump beams. Among these, the $\nu_{23}$ component only interacts with atoms with near-zero transverse velocity. For atoms with certain non-zero velocities, such as those in the $+\upsilon$ velocity group that satisfy the Doppler condition $\nu_{-1}+\vec{k_1}\cdot\vec{\upsilon}=\nu_{23}$, the pump beam from the $\nu_{-1}$ component (Pump2, see the bottom of Fig. \ref{fig1}a) excites the $F$=2 $\rightarrow$ $F’$=3 transition, where $\vec{k_1}$ is the wave vector of the Pump2 laser. Similarly, the $+\upsilon$ velocity group also satisfies the Doppler condition $\nu_{+1}-\vec{k_2}\cdot\vec{\upsilon}=\nu_{23}$, allowing the probe beam from the $\nu_{+1}$ component (Probe1) to excite the $F$=2 $\rightarrow$ $F’$=3 transition, with $\vec{k_2}$ as the wave vector of the Probe1 laser. Both Pump2 and Probe1 interact with atoms in the $+\upsilon$ velocity group, thereby enhancing the spectral amplitude of the $F$=2 $\rightarrow$ $F’$=3 transition. For the same reason, Pump1 and Probe2 can excite atoms in the $-\upsilon$ velocity group to participate in the $F$=2 $\rightarrow$ $F’$=3 transition. Additionally, minor contributions arise from the weaker $\pm$2nd-order sidebands ($\Delta'$ = 100 MHz) and $\pm$3rd-order sidebands ($\Delta''$ = 150 MHz), which are not shown in Fig. \ref{fig1}a. These atoms excited by the multi-frequency components form a comb-like distribution in the transverse velocity domain, ultimately resulting in an accumulated enhancement in the amplitude of the velocity-comb atomic spectroscopy.

The spectra before and after modulation by the EOM1, as measured by a spectrum analyzer, are shown in Fig. \ref{fig1}b, with the ordinate scale being linear in intensity. According to the characteristics of the Bessel function, within a certain modulation depth, the sideband intensity of EOM1 phase modulation decreases significantly with increasing order. Therefore, we select an appropriate modulation amplitude to make the intensity of the $\pm$1st-order sidebands comparable to the main frequency, while the $\pm$2nd and $\pm$3rd-order sidebands play a weaker role. When the single-frequency intensity is consistent with the triple-frequency (0 and $\pm$1st-order) intensities, we measure the SAS and MTS signals for both lasers and present the results in Fig. \ref{fig1}c. The left and right sides correspond to the $^{85}$Rb 5$^2$S$_{1/2}$ $F$=3 and $^{87}$Rb 5$^2$S$_{1/2}$ $F$=2 to 5$^2$P$_{3/2}$ transitions, respectively. It is evident that the SAS and MTS signals of the triple-frequency laser exhibit a multi-peak structure with increased atomic interactions, and the amplitude is significantly enhanced compared to that of the single-frequency laser. To quantitatively demonstrate the enhancement of the spectral signal, we then measure the SAS and MTS under different numbers of frequency components and calibrate the MTS amplitude of the $^{87}$Rb 5$^2$S$_{1/2}$ $F$=2 $\rightarrow$ 5$^2$P$_{3/2}$ $F’$=3 cycling transition.

\begin{figure*}[htbp]
\centering
\captionsetup{singlelinecheck=no, justification = RaggedRight}
\includegraphics[width=\textwidth]{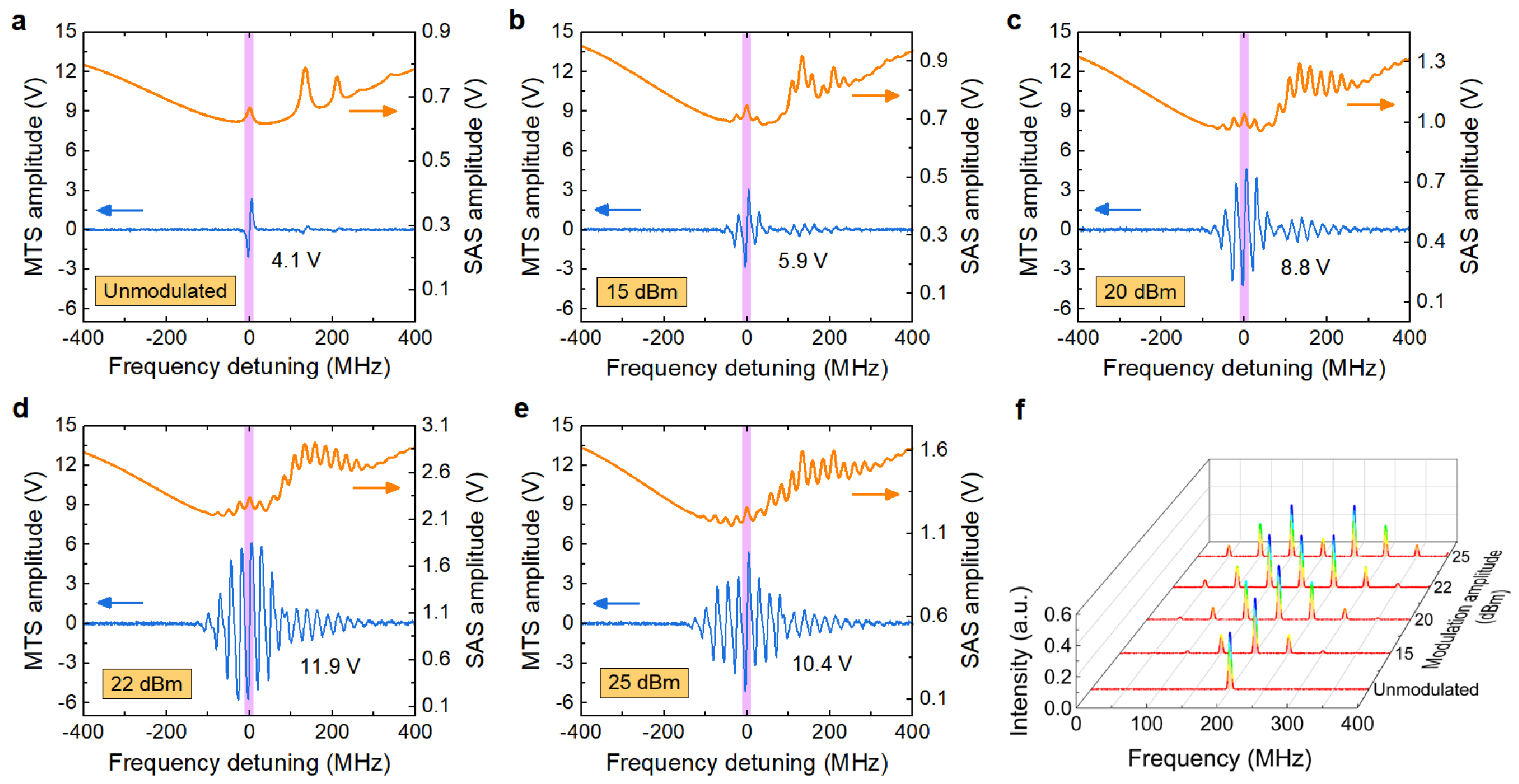}
\caption{\textbf{Velocity-comb atomic spectroscopy under different frequency components} (orange curve: SAS, blue curve: MTS)\textbf{.} From \textbf{a} to \textbf{e}, the modulation amplitudes of EOM1 are: unmodulated, 15 dBm, 20 dBm, 22 dBm, and 25 dBm, respectively. The purple-shaded region corresponds to the $^{87}$Rb 5$^2$S$_{1/2}$ $F$=2 $\rightarrow$ 5$^2$P$_{3/2}$ $F’$=3 cycling transition. \textbf{f} Laser spectra at different modulation amplitudes.   
}
\label{fig2}
\end{figure*}

\vspace{3pt}
\noindent 
\maketitle
\textbf{Relationship between frequency comb components and MTS amplitude}\\
\noindent 
The number of frequency components can be controlled by adjusting the modulation amplitude of EOM1 (see Fig. \ref{fig2}f), but the peak intensity of the main components needs to be kept equal. When the laser is unmodulated, it has only one frequency component, resulting in the traditional single-frequency spectrum shown in Fig. \ref{fig2}a, with a corresponding MTS amplitude of 4.1 V. When the modulation amplitude is 15 dBm, two small 1st-order sidebands appear on either side of the main frequency. Although their intensities are relatively weak, we can still distinguish the multi-frequency effect and the increase in MTS amplitude from Fig. \ref{fig2}b. When the modulation amplitude is 20 dBm, the $\pm$1st-order sidebands become more prominent, and the enhancement of the velocity-comb atomic spectroscopy can be clearly seen in Fig. \ref{fig2}c, with a corresponding MTS amplitude of 8.8 V. Further increasing the modulation amplitude to make the intensities of the $\pm$1st-order sidebands equal to the main frequency. At 22 dBm, we obtain a triple-frequency spectrum as shown in Fig. \ref{fig2}d. The multi-peak structure is more pronounced, and the MTS amplitude reaches 11.9 V. When the modulation amplitude is further increased to 25 dBm, the enhanced sidebands begin suppressing the main frequency, and the $\pm$1st-order sidebands dominate the contribution. As shown in Fig. \ref{fig2}e, the amplitude of the dual-frequency spectrum is slightly lower than that of the triple-frequency spectrum, and the corresponding MTS signal is reduced to 10.4 V. From these comparative results, we can infer that the enhancement of spectral amplitude is positively correlated with the number of laser frequency components. However, it should be noted that each frequency component requires a certain power allocation. In this work, the total MTS powers corresponding to the five cases of unmodulated, 15 dBm, 20 dBm, 22 dBm, and 25 dBm are 0.17 mW, 0.2 mW, 0.28 mW, 0.61 mW, and 0.36 mW, respectively.

\vspace{3pt}
\noindent 
\maketitle
\textbf{Characterization of velocity-comb MTS}\\
\noindent
Under the modulation amplitude of 22 dBm, we further investigate the working characteristics of the velocity-comb MTS. Figure \ref{fig3}a shows the relationship between the MTS slope and the pump laser power, corresponding to the $^{87}$Rb 5$^2$S$_{1/2}$ $F$=2 $\rightarrow$ 5$^2$P$_{3/2}$ $F’$=3 cycling transition (same below). The red, orange, and blue points represent measurements taken at probe laser powers of 0.05 mW, 0.1 mW, and 0.15 mW, respectively. As the pump laser power increases, all three curves exhibit an initial increase followed by a decrease, with the MTS slope reaching its maximum near 0.4 mW. Figure \ref{fig3}b depicts the relationship between the MTS slope and the probe laser power at a fixed pump laser power of 0.4 mW. When the probe laser power is approximately 0.15 mW, the MTS slope approaches saturation. Moreover, the velocity-comb atomic spectroscopy is also influenced by the temperature of the Rb cell. As the temperature rises, the increasing atomic density leads to an initial rise in the MTS slope (see Fig. \ref{fig3}c). However, once the temperature exceeds a certain value (50$^\circ$C), intensified atomic collision broadening limits further increases in the slope.

\begin{figure}[htbp]
\centering
\captionsetup{singlelinecheck=no, justification = RaggedRight}
\includegraphics[width=0.48\textwidth]{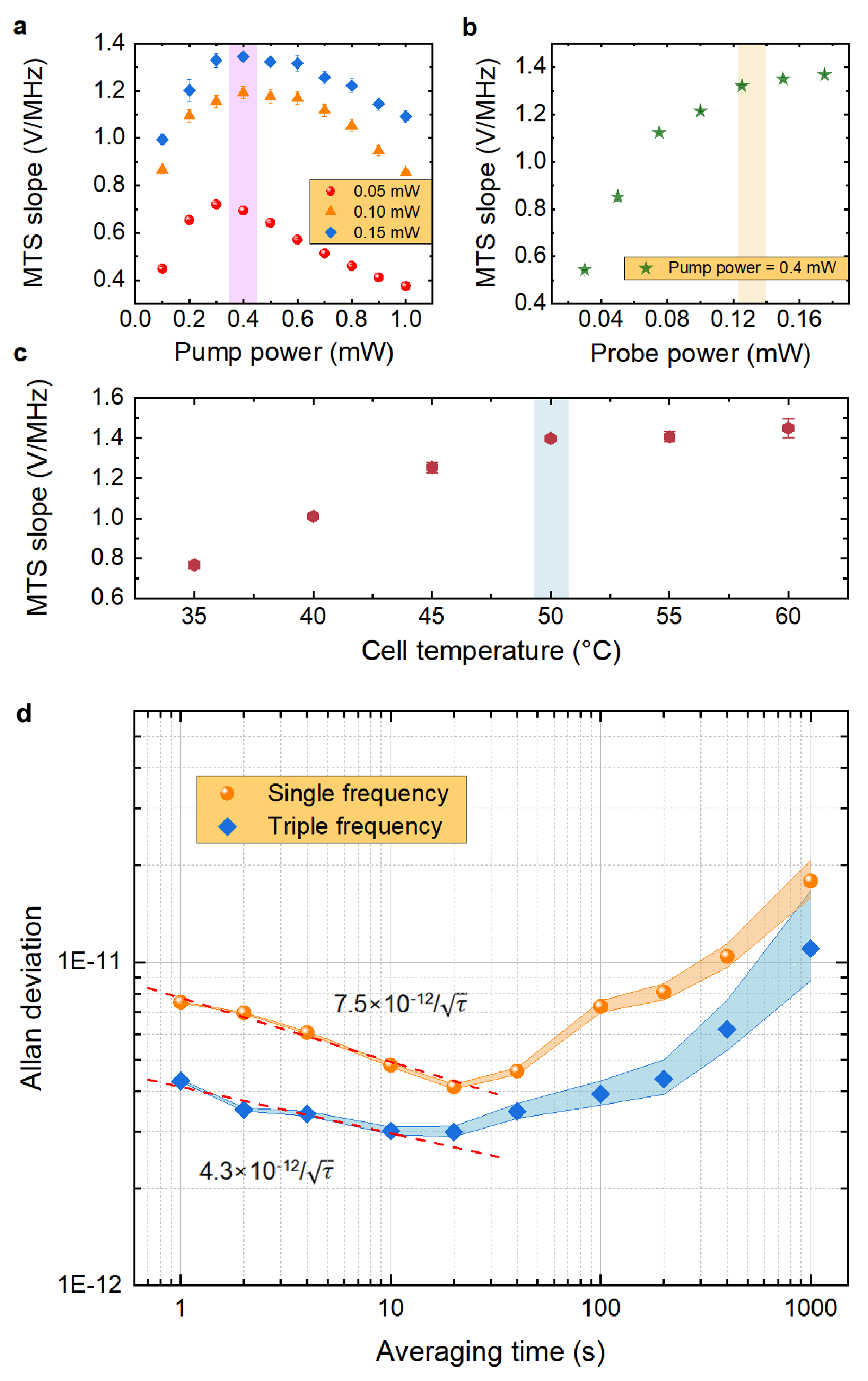}
\caption{\textbf{Working characteristics of velocity-comb MTS.} The slope of the velocity-comb MTS as a function of \textbf{a} pump laser power, \textbf{b} probe laser power, and \textbf{c} cell temperature. The red, orange, and blue points in \textbf{a} correspond to probe laser powers of 0.05 mW, 0.1 mW, and 0.15 mW, respectively. All values are measured for the $^{87}$Rb 5$^2$S$_{1/2}$ $F$=2 $\rightarrow$ 5$^2$P$_{3/2}$ $F’$=3 cycling transition. \textbf{d} Allan deviations of the single-frequency laser (orange) and the triple-frequency laser (blue) stabilized by MTS. The shaded areas denote the error bands. The red dashed lines represent the linear fit to the short-term data points.   
}
\label{fig3}
\end{figure}

\vspace{3pt}
\noindent 
\maketitle
\textbf{Verification of frequency stability improvement}\\
\noindent
Next, we experimentally verify the improvement in frequency stability achieved with the velocity-comb MTS scheme. The relationship between the number of frequency components and the spectral $S/N$ has been derived in detail in Ref. \cite{40}, so we only list a few key formulas here. The saturated absorption signal and the spectral $S/N$ of a single-frequency laser follow the expressions:
\begin{equation}
P_S(\omega)=P\left[1-\kappa\frac{(\Gamma/2)^2}{(\omega-\omega_0)^2+(\Gamma/2)^2}\right]
\end{equation}
and
\begin{equation}
\frac{S}{N}_{(S)}=\kappa\sqrt{\frac{P\tau}{\hbar\omega}},
\end{equation}
where $\omega$ is the laser frequency, $P$ is the laser power, $\kappa$ is the dimensionless saturation parameter, $\omega_0$ is the center frequency of the atomic transition, $\Gamma$ is the homogeneous linewidth, and $\tau$ is the duration. When a multi-frequency laser is applied to the atoms, each pair of oppositely directed pump and probe lasers can interact with atoms from different transverse velocity groups and independently contribute to the absorption coefficient. Thus, the saturated absorption signal of the multi-frequency laser can be expressed as:
\begin{equation}
P_M(\omega)=\sum_{n}P_n\left[1-\kappa_n\frac{(\Gamma/2)^2}{(\omega-\omega_0)^2+(\Gamma/2)^2}\right].
\end{equation}
For $N$ frequency components, the spectral $S/N$ is then deduced as:
\begin{equation}
\frac{S}{N}_{(M)}=\kappa\sqrt{\frac{NP\tau}{\hbar\omega}},
\end{equation}
which is improved by a factor of $\sqrt{N}$ compared to the single-frequency spectrum.

Stabilizing the frequency of the 780 nm ECDL to the $F$=2 $\rightarrow$ $F’$=3 cycling transition of the single-frequency spectrum shown in Fig. \ref{fig2}a and the triple-frequency spectrum shown in Fig. \ref{fig2}d, respectively, and using an ultra-stable laser with known high frequency stability ($10^{-13}$ level, compared with an optical frequency comb) to perform beat measurements with both, we obtain the frequency stability comparison results shown in Fig. \ref{fig3}d. Over an averaging time of 10 s, the Allan deviations of the single-frequency laser and the triple-frequency laser are 7.5 $\times$ $10^{-12}$/$\sqrt{\tau}$ and 4.3 $\times$ $10^{-12}$/$\sqrt{\tau}$, respectively. Across the entire averaging time measurement range, the frequency stability of the triple-frequency laser is improved by nearly $\sqrt{3}$ times compared to the single-frequency laser. This result is in good agreement with the theoretical expectation. As a preliminary proof of principle, the optical path is currently dispersed on the optical table, without actively controlling environmental factors such as temperature and mechanical vibrations. Therefore, the system's frequency stability is limited to the $10^{-12}$ level. Upon validation of this scheme, we will integrate the system with anti-interference measures and optimize residual amplitude modulation (RAM) \cite{41,42,43}, power stability \cite{44}, etc., aiming to further improve frequency stability based on the best-performing thermal vapor-cell rubidium optical frequency standard \cite{34}. In addition, we find that by locking the laser frequency to the zero-crossing points of different dispersion curves in the velocity-comb MTS, various stabilized lasers close to the atomic transition frequency can be easily achieved, which is an additional advantage of this scheme.

\vspace{3pt}
\noindent 
\maketitle
\begin{figure}[htbp]
\centering
\captionsetup{singlelinecheck=no, justification = RaggedRight}
\includegraphics[width=0.47\textwidth]{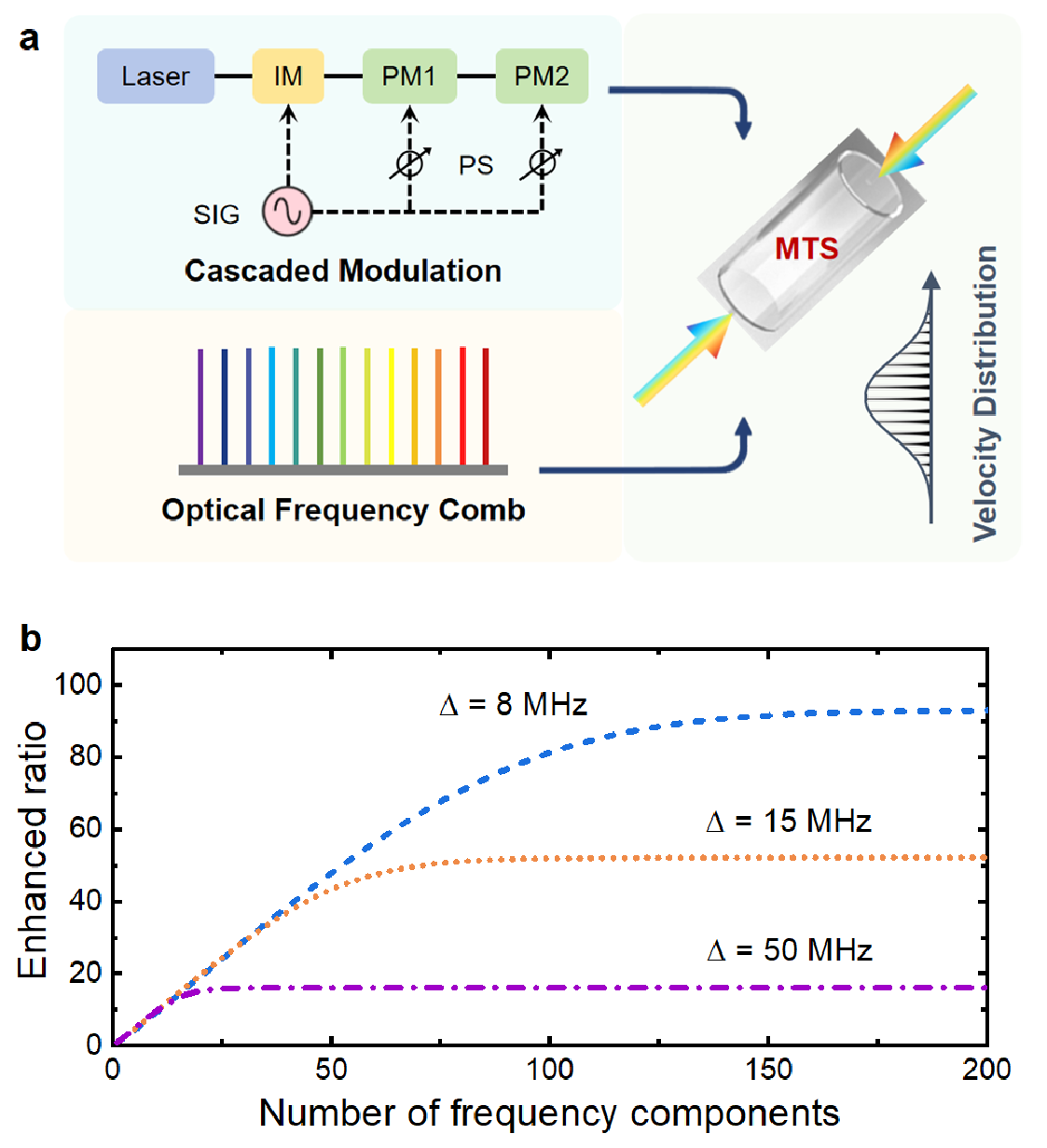}
\caption{\textbf{Prospects of velocity-comb MTS.} \textbf{a} Increased frequency components for MTS stabilization using cascaded modulation or an optical frequency comb. IM: intensity modulator; PM: phase modulator; SIG: signal generator; PS: phase shifter. \textbf{b} Amplitude enhancement ratio of the velocity-comb atomic spectroscopy as a function of the number of frequency components ($N$) for different frequency intervals ($\Delta$).
}
\label{fig5}
\end{figure}

\textbf{Prediction for $10^{-15}$ stability lasers referenced to thermal Rb atoms}\\
\noindent
Among alkali metals, Rb stands out for its low melting point, high abundance, and ease of extraction. A thermal atomic vapor cell heated above room temperature can produce sufficient saturated vapor pressure, making Rb an attractive quantum reference ensemble for low-cost and miniaturized optical clocks. Over decades of technological advancement, MTS stabilization based on Rb atoms has commonly achieved laser frequency stabilities on the order of $10^{-14}$ \cite{34,69,65}. Following the approach proposed in this work, we estimate that the frequency stability of an Rb 780 nm laser could potentially reach the $10^{-15}$ level. The specific scheme is illustrated in Fig. \ref{fig5}a.

Using cascaded modulation \cite{50,51} or an optical frequency comb \cite{52} to generate multiple frequency comb components and applying them to the MTS stabilization system, we theoretically calculate the amplitude enhancement ratio of the velocity-comb atomic spectroscopy under different frequency intervals ($\Delta$ = 8 MHz, 15 MHz, and 50 MHz) and numbers of frequency components ($N$). The ratio is simply given by:
\begin{equation}
\mathcal{R} \propto \frac{\displaystyle\sum_{n}(e^{-\alpha_{n}^{'}l}-e^{-\alpha_{0}^{'}l})}{e^{-\alpha l}-e^{-\alpha_{0}l}},
\end{equation}
where $\alpha$ and $\alpha_0$ represent the absorption coefficients of the probe laser in the single-frequency case, with and without the pump laser, respectively; $\alpha_{n}^{'}$ and $\alpha_{0}^{'}$ represent the absorption coefficients of the probe laser corresponding to the $n$-th frequency in the multi-frequency comb case, with and without the pump laser, respectively; $l$ is the length of the Rb cell. As shown in Fig. \ref{fig5}b, for a fixed $\Delta$, the ratio initially increases linearly with $N$ but gradually saturates once the total spectral width exceeds the Doppler range. Smaller $\Delta$ values allow more frequency components within the Doppler range, significantly increasing the amplitude enhancement ratio by engaging more atomic velocity groups in the transition. When $\Delta$ = 8 MHz, the ratio approaches 100-fold. For the 420 nm transition (5$^2$S$_{1/2}$ $\rightarrow$ 6$^2$P$_{3/2}$, $\Gamma$=1.45 MHz), which has a narrower natural linewidth compared to the 780 nm transition (5$^2$S$_{1/2}$ $\rightarrow$ 5$^2$P$_{3/2}$, $\Gamma$=6.06 MHz), we predict that it holds great potential for achieving order-of-magnitude improvement in laser frequency stability \cite{65}.

\vspace{6pt}
\noindent
\large\textbf{Discussion} \\
\normalsize
\noindent 
In this work, we experimentally demonstrate a velocity-comb MTS system. By applying external modulation to the single-frequency laser, many atoms with non-zero transverse velocities can be excited by the multi-frequency comb components and participate in the transition process, significantly increasing the $S/N$ of the spectrum. Experimental results show that the frequency stability of the triple-frequency laser is nearly $\sqrt{3}$ times greater than that of the single-frequency laser, successfully confirming the initial conception of this scheme. In future work, we plan to add cascaded modulation \cite{50,51} or use optical frequency combs \cite{52} to generate more frequency components, aiming for achieving a frequency stability of $10^{-15}$ for the thermal vapor-cell rubidium optical frequency standard and $10^{-16}$ for the iodine molecular optical frequency standard. Furthermore, this method is universal for any quantum system with a wide velocity distribution. It can also be incorporated into existing textbooks \cite{1} as an emerging spectroscopic technique, potentially opening up broader application scenarios in fundamental science \cite{45,46,47} and advanced technology \cite{48,49} through continuous advancements.

\vspace{6pt}
\noindent \textbf{Methods}\label{Methods}\\
\begin{footnotesize}
\noindent \textbf{Experimental details.} 
The experimental setup of the velocity-comb MTS system is illustrated in Fig. \ref{fig4}. The output laser from the 780 nm ECDL first passes through an isolator (ISO) to suppress back reflections. It is then split into two beams by a half-wave plate (HWP) and a polarizing beam splitter (PBS), with the transmitted beam serving as the output light and the reflected beam directed for multi-frequency modulation. The EOM1, with a resonance frequency of 50 MHz mentioned above, is used to modulate the single-frequency laser into a multi-frequency beam. To observe the frequency distribution of the multi-frequency laser, the single-frequency laser is shifted by 200 MHz using an acousto-optic modulator (AOM) before EOM1 modulation. By combining the frequency-shifted laser with the multi-frequency laser and performing heterodyne measurements, the complete spectrum is displayed on the spectrum analyzer (Keysight N9010B). 

\begin{figure}[htbp]
\centering
\captionsetup{singlelinecheck=no, justification = RaggedRight}
\includegraphics[width=0.49\textwidth]{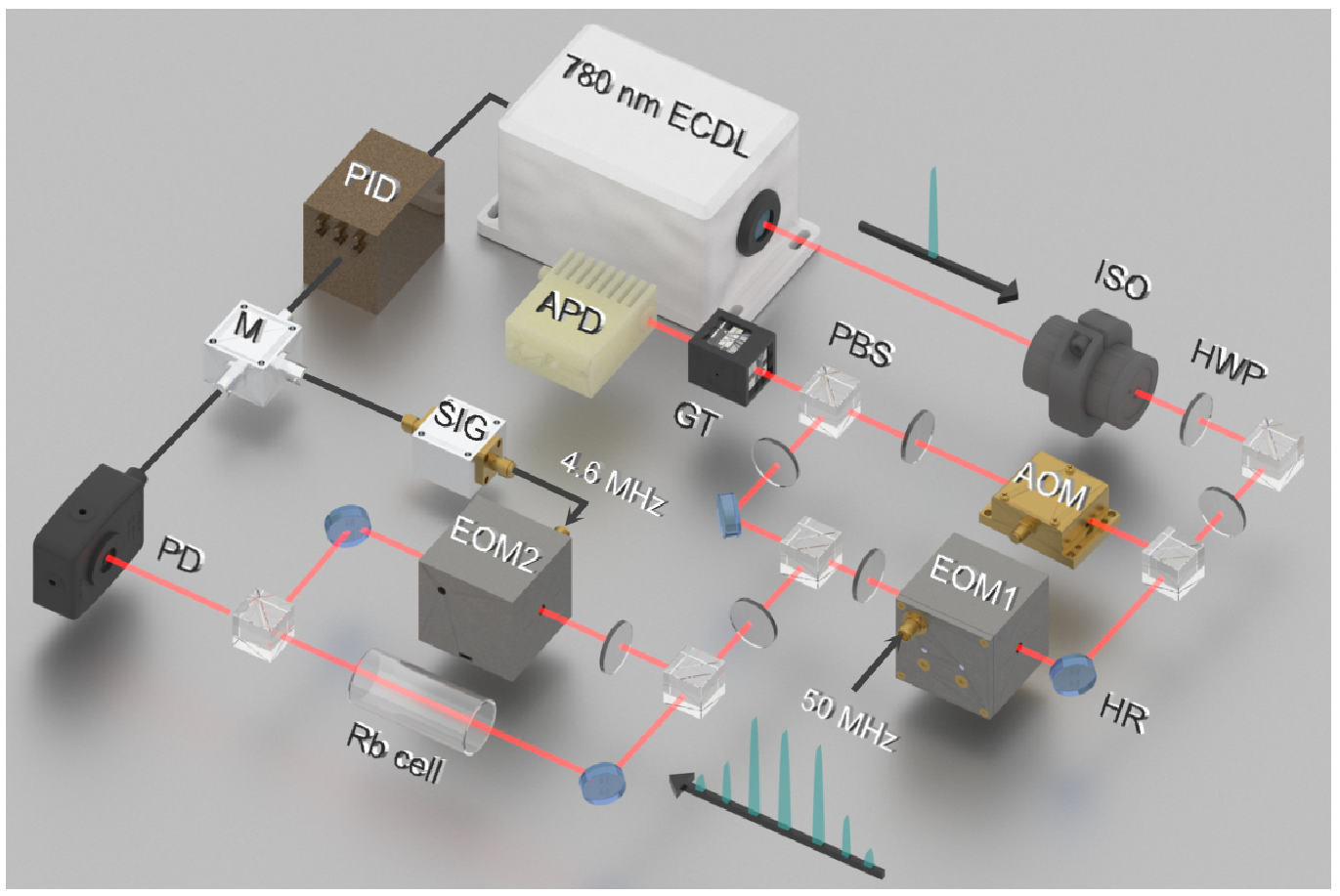}
\caption{\textbf{Experimental setup of the velocity-comb MTS system.} ECDL: external cavity diode laser; ISO: isolator; HWP: half-wave plate; PBS: polarizing beam splitter; HR: high-reflection mirror; EOM: electro-optic modulator; AOM: acousto-optic modulator; GT: Glan-Taylor prism; APD: avalanche photodiode; PD: photo-electric detector; SIG: signal generator; M: mixer; PID: proportion-integral-derivative locking system.   
}
\label{fig4}
\end{figure}

The multi-frequency laser output from EOM1 then enters the MTS system for frequency stabilization. It is further divided into probe and pump beams. The pump beam is phase-modulated by an electro-optic modulator (EOM2) with a resonance frequency of 4.6 MHz. After reflection, it passes through the Rb cell in reverse coincidence with the probe beam. The HWP placed before EOM2 is used to adjust the laser’s polarization direction, aligning it with the main axis of the EOM2 crystal. The Rb cell is temperature-controlled and magnetically shielded to isolate it from environmental interference. After a four-wave mixing process with the pump beam in the Rb cell, the probe beam is received by a photo-electric detector (PD), which contains modulation information. The signal received by PD is filtered, amplified, and then mixed with a sine signal at the same frequency as the EOM2 driving signal. The dispersion-like signal (velocity-comb MTS) output from the mixer serves as the error signal, which is fed back to three different frequency response ports through a proportion-integral-derivative (PID) locking system: slow PZT port, slow current port, and fast diode current port. By locking the frequency of the 780 nm ECDL to the zero-crossing point of the error signal, a stabilized laser corresponding to the hyperfine transition frequency of Rb atoms is achieved.

\vspace{6pt}
\noindent\textbf{Data availability}\\
The data that supports the plots within this paper and other findings of this study are available from the corresponding authors upon reasonable request. 

\end{footnotesize}
\vspace{20pt}

\bibliography{Ref.bib}


\vspace{6pt}
\begin{footnotesize}

\vspace{6pt}
\noindent \textbf{Acknowledgment}

\noindent 
This research was funded by the Innovation Program for Quantum Science and Technology (2021ZD0303200), China Postdoctoral Science Foundation (BX2021020), National Natural Science Foundation of China (NSFC) (62405007), Science and Technology on Metrology and Calibration Laboratory (JLJK2023001B001), and Natural Science Foundation of Jiangsu Province (BK20240613).

\vspace{6pt}
\noindent \textbf{Author contributions}

\noindent 
J.C. conceived the idea of velocity-comb modulation transfer spectroscopy by improving the atomic utilization ratio in frequency-stabilized laser systems. X.G. performed the experiments and wrote the manuscript. X.G. and Z.X. carried out the theoretical calculations. Z.L., Z.W., J.Z., Y.W., P.C. and T.S. participated in the discussion of results and revision of the manuscript.

\vspace{6pt}
\noindent \textbf{Competing interests} 

\noindent The authors declare no competing interests.
\end{footnotesize}

\end{document}